\documentclass[draft,unnumberedheadings,letterpaper]{aipproc}

\layoutstyle{8x11single}
\newcommand{\lya}{Ly$\alpha$}

\usepackage[]{pslatex}
\usepackage[]{natbib}
\usepackage[]{amssymb}

\usepackage[]{myusepack}

\usepackage[]{floatflt}

\begin{document}

{\noindent Submitted in response to a 
Request for Information: Science Objectives and Requirements for the Next NASA UV/Visible Astrophysics Mission Concepts \\[.2in] }
\noindent Primary Contact: Stephan R. McCandliss, Department of Physics and Astronomy, The Johns Hopkins University, Baltimore, MD  21218, tel 410-516-5272; stephan@pha.jhu.edu

\title[Project Lyman]{Project Lyman: Quantifying 11 Gyrs of Metagalactic Ionizing Background Evolution }


\author{Stephan R. McCandliss~(jhu.edu),  B-G Andersson~(usra.edu), Nils Bergvall~(uu.se), Luciana Bianchi~(jhu.edu), Carrie Bridge~(caltech.edu), Milan Bogosavljevic~(caltech.edu), Seth H. Cohen~(asu.edu), Jean-Michel Deharveng~(oamp.fr), W. Van Dyke Dixon (jhu.edu), Harry Ferguson~(stsci.edu), Peter Friedman~(caltech.edu), Matthew Hayes~(unige.ch), J. Christopher Howk~(nd.edu) Akio Inoue~(osaka-sandai.ac.jp), Ikuru Iwata~(nao.ac.jp), Mary Elizabeth Kaiser~(jhu.edu), Gerard Kriss~(stsci.edu), Jeffrey Kruk~(nasa.gov), Alexander S. Kutyrev~(gsfc.nasa.gov), Claus Leitherer~(stsci.edu), Gerhardt R. Meurer~(uwa.edu.au), Jason X. Prochaska~(ucolick.edu), George Sonneborn~(gsfc.nasa.gov), Massimo Stiavelli~(stsci.edu), Harry I. Teplitz~(caltech.edu), Rogier A Windhorst~(asu.edu)}{address={}, altaddress={}}

\begin{abstract}

The timing and duration of the reionization epoch is crucial to the emergence and evolution of structure in the universe.  The relative roles that star-forming galaxies, active galactic nuclei and quasars play in contributing to the metagalactic ionizing background across cosmic time remains uncertain.  Deep quasar counts provide insights into their role, but the potentially crucial contribution from star-formation is highly uncertain due to our poor understanding of the processes that allow ionizing radiation to escape into the intergalactic medium (IGM).  The fraction of ionizing photons that escape from star-forming galaxies is a fundamental free parameter used in models to "fine-tune" the timing and duration of the reionization epoch that occurred somewhere between 13.4 and 12.7 Gyrs ago (redshifts between  12 > z > 6). However, direct observation of Lyman continuum (LyC) photons emitted below the rest frame \ion{H}{1} ionization edge at 912  \AA\ is increasingly improbable at redshifts $z >$ 3, due to the steady increase of intervening Lyman limit systems towards high $z$.  

Thus UV and U-band optical bandpasses provide the only hope for direct, up close and in depth, observations of the types of environment that favor LyC escape.  By quantifying the evolution over the past 11 billion years ($z < $3) of the relationships between LyC escape and local and global parameters such as: metallicity, gas fraction, dust content, star formation history, mass, luminosity, redshift, over-density and quasar proximity,  we can provide definitive information on the LyC escape fraction that is so crucial to answering the question of, how did the universe come to be ionized?  Here we provide estimates of the ionizing continuum flux emitted by "characteristic" ($L_{uv}^*$) star-forming galaxies as a function of look back time and escape fraction, finding that at $z =$ 1 (7.6 Gyrs ago) $L_{uv}^*$ galaxies with an escape fraction of 1\% have a flux of 10$^{-19}$ ergs cm$^{-2}$ s$^{-1}$ \AA$^{-1}$.

\end{abstract}

\setcounter{page}{0}
\thispagestyle{empty}
\maketitle


\setcounter{page}{1}


\section{The essential role of LyC escape  }

Some 0.3 Myr after the Big Bang, the adiabatic expansion of the universe caused the primordial plasma of protons and electrons to cool, creating a neutral gas.
Recent observations show that most of the universe has since been reionized and provide constraints to the duration of this process.  Sloan Digital Sky Survey spectra of luminous high-redshift quasars have black \ion{H}{1} Gunn-Peterson troughs, indicating a mean \ion{H}{1} fraction of $\gtrsim 10^{-3}$ at  $z \ge $ 6.4  when the age of universe was $\approx$ 1 Gyr \citep{Fan:2006}.  Evidence that reionzation started even earlier is provided by the polarization of the microwave background on large angular scales seen by the Wilkinson Microwave Anistropy Probe, which is consistent with an ionization fraction $\sim$ unity at $z \approx$ 11 when the universe was $\approx$ 365 Myr old \citep{Spergel:2007}.  

Complete reionization occurs when the rate of ionizing photons emitted within a recombination time exceeds the number of neutral hydrogen atoms.  The duration of the reionization epoch depends on the initial mass function (IMF) of the first ionizing sources, their intrinsic photoionization rate ($Q$), the baryon clumping factor ($C\equiv {<\rho^2>/<\rho>^2}$), and the fraction of ionizing photons that somehow manage to escape into the intergalactic medium (IGM)  \citep{Madau:1999}.  Of these parameters the LyC escape fraction ($f_e$) is the  least constrained  \citep{Fan:2006}. Its (often arbitrary) choice can alter conclusions regarding the nature and duty cycle of the sources thought to be responsible for initiating and sustaining  reionization \citep[]{Gnedin:2000,Fernandez:2011}. 

LyC escape plays an essential role in the formation of structure.  The escape fraction of ionizing photons from galaxies is the single greatest uncertainty in estimating the intensity of the metagalactic ionizing background (MIB) over time \citep{Heckman:2001}. The MIB controls the ionization state of the IGM at all epochs and may be responsible for hiding a non-trivial fraction of the baryons in the universe \citep[c.f.][]{Tripp:2008, Danforth:2008}.  Ionizing radiation produced by star-forming galaxies is ultimately related to the rate of metal production by stellar nucleosynthesis \citep{Gnedin:1997}. The MIB intensity is a gauge of the feedback into the IGM of chemicals, mechanical energy and radiation by supernovae and stellar winds.

\begin{figure}
\includegraphics*[scale=.25]{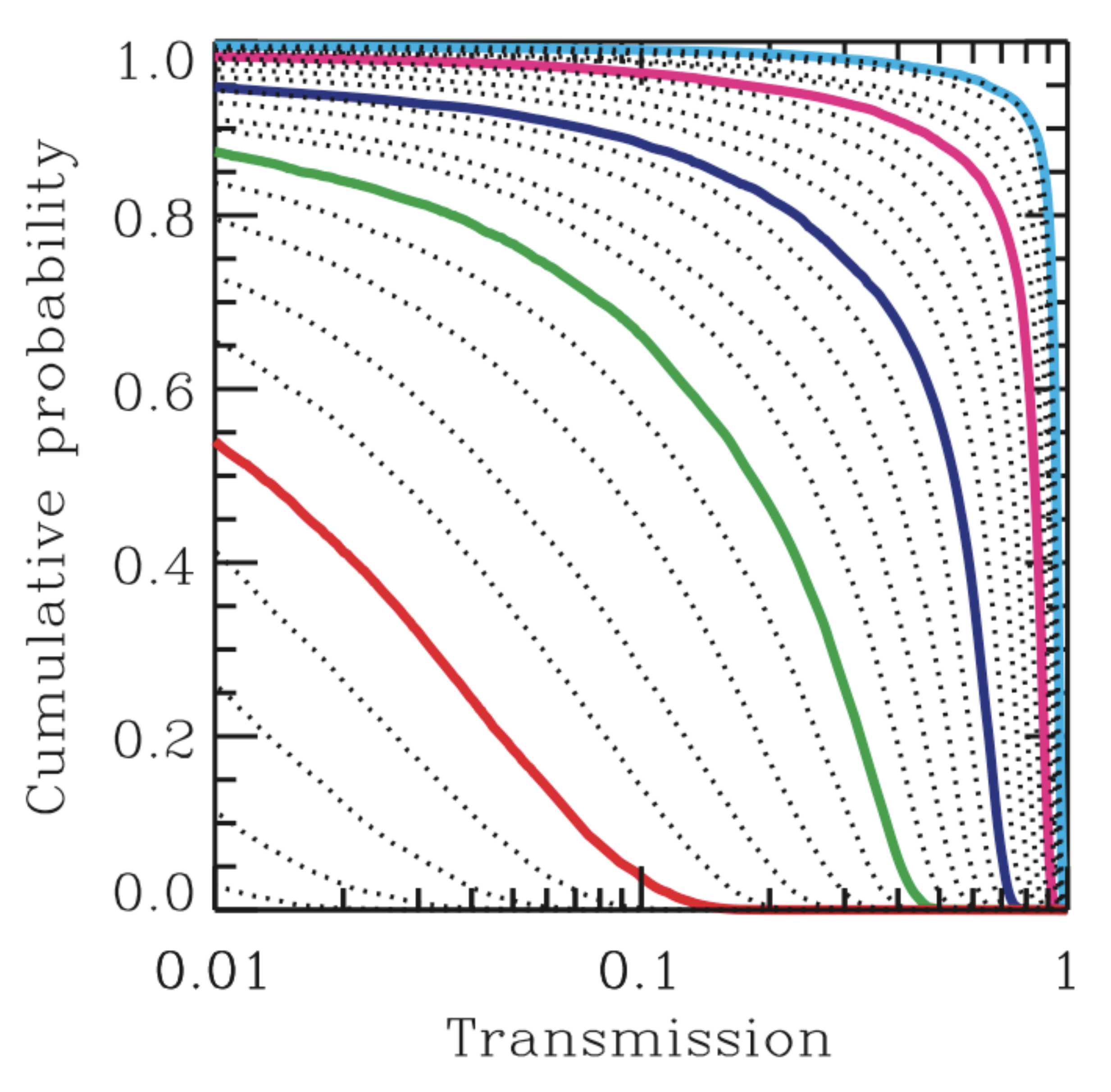} 
\parbox[b]{3in}{ \small {\textbf{\textsc{figure~1.} } } The results of a Monte Carlo \citep{Inoue:2008} showing the effects of intergalactic absorption by Lyman limit systems on the transmission of LyC photons.  The graph depicts the cumulative probability of having a line-of-sight transmission greater than that shown on the axis for LyC photons emitted between (880 -- 910 \AA).  The light blue, magenta, dark blue, green and red are contours for the redshifts $z = $ 1, 2, 3, 4 and 5 respectively.  At $z =$ 4, the probability of having a transmission of greater than 30\% is 0.2 \label{inoue} \vspace*{.75in}}. 
\end{figure}
\setcounter{figure}{1}

Ionizing radiation regulates the collapse of baryons on local and global scales \citep{Ricotti:2008}.  Photoelectrons provide positive feedback for star formation by promoting the formation of H$^{-}$, which in turn catalyzes the production of H$_{2}$; a crucial coolant for collapse at high-z.  Negative feedback occurs when photoionization heating raises the temperature and inhibits stellar collapse by increasing the Jeans mass.  Photodissociation of H$_{2}$ is another form of negative radiative feedback mediated by both Lyman-Werner photons in the 912 -- 1120 \AA\ bandpass and LyC photons \citep{McCandliss:2007}.  Whether ionizing radiation has a positive or negative effect on a collapsing body depends on the gas density, the strength of the radiation field, the source lifetime  and  the escape fraction of LyC photons\citep{Ricotti:2008, Ciardi:2008}. 

Quantification of the LyC escape fraction is at the frontier of reionization physics.  The high opacity of even small column densities of \ion{H}{1} to ionizing radiation makes the sources very faint at all epochs, but especially at redshifts $z \gtrsim$ 3 \citep{Madau:1999}.  Figure~1 from \citet{Inoue:2008}, shows the likelihood of detecting LyC escape from star-forming galaxies becomes increasingly improbable above $z >$ 3, due to a progressive increase with redshift in the number density of intervening Lyman limit and \lya\ forest systems.  Detections above $z >$ 4, while not ruled out, will be extremely rare.  This favors UV and U-band optical observations in efforts to directly identify and spatially resolve those physical environments that allow LyC to escape.  By examining at low redshift the relationship between LyC escape and the local and global parameters of metallicity, gas fraction, dust content, star formation history, mass, luminosity, redshift, over-density and quasar proximity, we seek to understand how the universe came to be ionized.

\section{Source(s) of reionization and the MIB} 
The fundamental question is, how did the universe come to be reionized and how long did it take?  Current thinking posits that LyC escape from the smallest galaxies powers reionization at $z \approx$ 6, since quasars are too few in number to sustain reionization \citep{Madau:1999, Bouwens:2008, Yan:2004}. 
However, this conclusion depends on an extrapolation of the faint end slope of the galaxy luminosity function (LF, -1.6 $\le \alpha \le $ -2.0), the faint end luminosity cutoff, the clumping factor (3 $ < C < $ 45) and escape fraction ($f_e \sim$ 0.1 -- 0.8).  

\begin{figure}[]  
\centering
\vspace*{-.5in}
\includegraphics[width=0.4\textwidth]{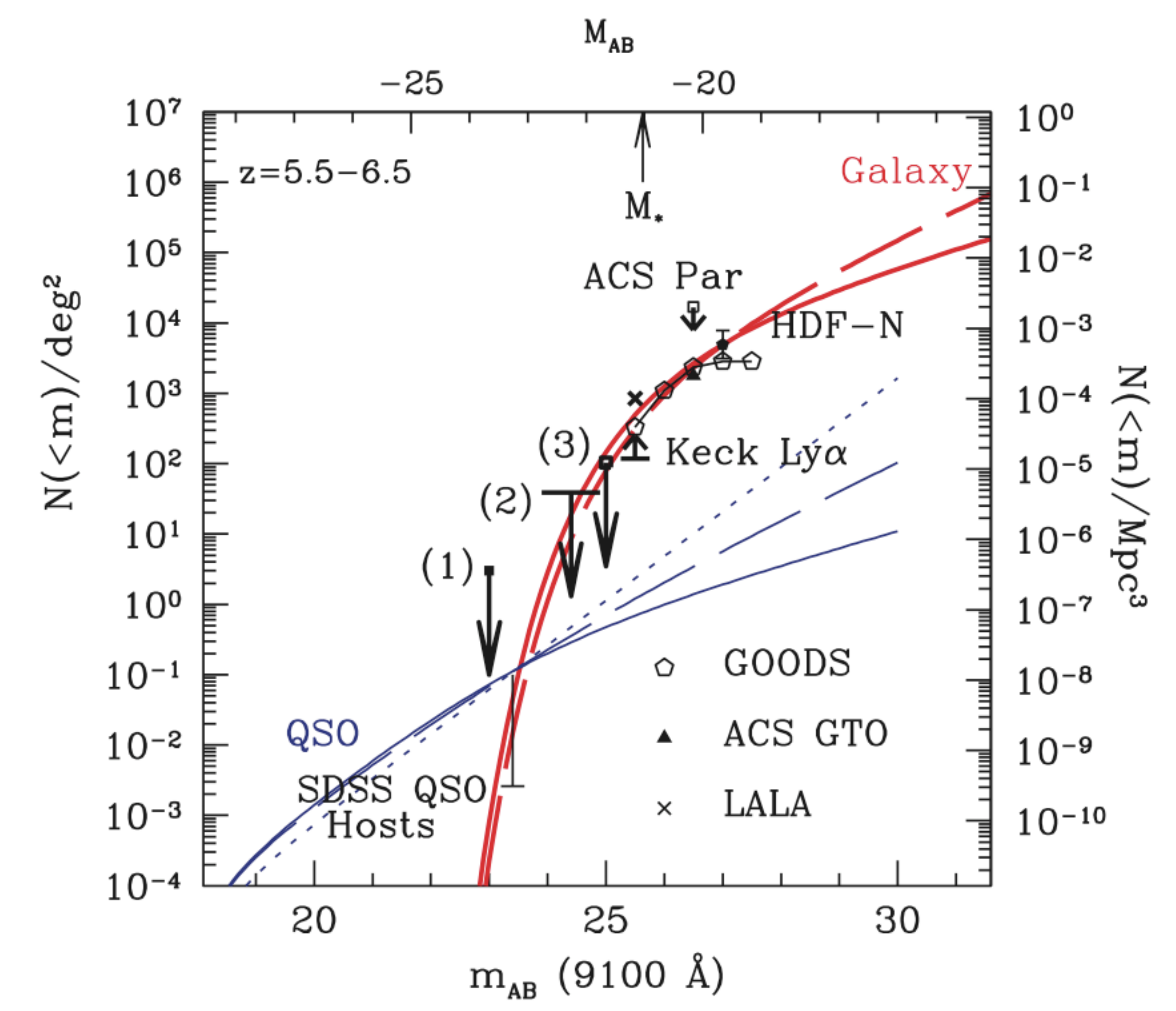}
\includegraphics[width=0.32\textwidth]{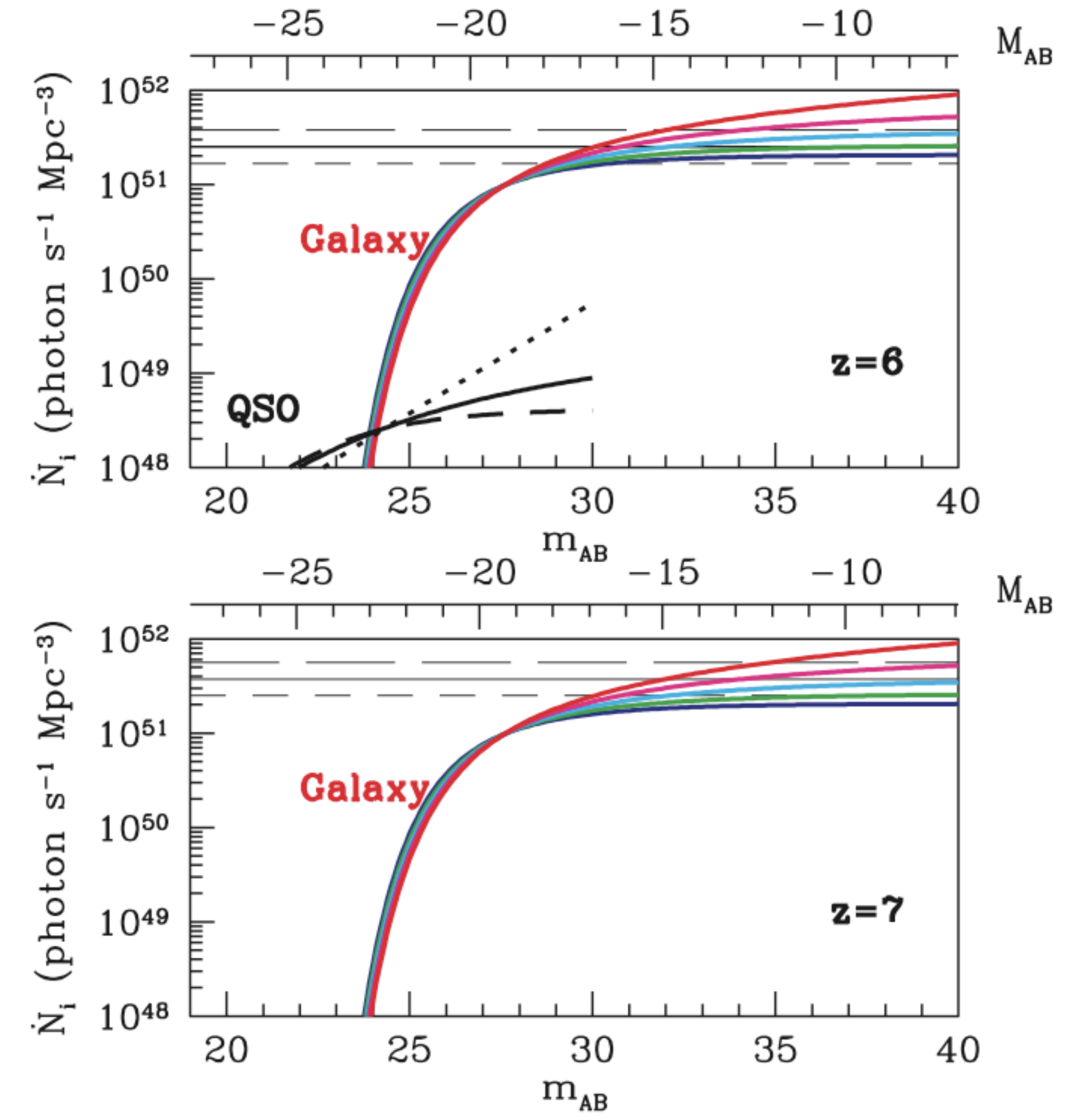}
\caption{Left --  LFs for quasars and galaxies at $z =$ 6.  Right -- Reionizing photon production rate for quasars and galaxies at a redshift of $z =$ 6 and for galaxies alone at  $z =$ 7.  The production rate required to maintain a fully ionized universe for clumping factors of 20, 30 and 40 are indicated at the top of each graph.  Higher clumping factors require more photon production. See \citep{Yan:2004} for details. }
\label{figyan}
\end{figure}

The sensitivity of this conclusion to the faint end slope and the role played by the clumping factor and $f_e$ is illustrated in Figure~\ref{figyan} taken from \citet{Yan:2004}.  On the left, the LFs for quasars and galaxies are displayed for a redshift of $z =$ 6.  Two extreme faint end slopes are shown for the galaxies (-1.6, -2.0) and for the quasars (-1.6, -2.6).  On the right, two panels show the cumulative reionizing photon production rate for quasars and galaxies at a redshift of $z =$ 6 and for galaxies alone at  $z =$ 7.  Horizontal lines drawn at the top of each panel mark the critical production rate required to keep the universe fully ionized for clumping factors of 20, 30, 45 (higher clumping factors require more photon production to overcoming clump self-shielding) and assuming $f_e =$ 0.1.

The figures show that at $z =$ 6 the faintest galaxies dominate the LyC production and are more likely than quasars to maintain the universe in a fully ionized state.  The case becomes less certain at $z =$ 7 where it has been found that maintaining reionization requires either a "top heavy" IMF or escape fractions 0.3 $\la f_e \la $ 0.8, assuming 20 $ < C < $ 45 \citep{Chary:2008, Meiksin:2005}, although recent work by \citet{Finkelstein:2010} suggests  0.1 $\la f_e \la $ 0.5 and 3 $ < C < $ 5.  It may be that there are not enough star-forming galaxies early on to initiate reionization \citep{Bolton:2007} and that mini-quasars might be involved \citep{Madau:2004}. There are also indications that the initiation of reionization above $z = $ 7 may require a "hard" spectral energy distribution (SED) more characteristic of quasars \citep{Bolton:2007}.  

Black holes reside in the nuclei of most if not all quiescent galaxies \citep{Ferrarese:2005, Soltan:1982}, so it is simplistic to characterize reionization as a process caused by either quasars with $f_e$ = 1 or star-forming galaxies with $f_e <$  1. Some fraction of quasars exhibit a break at the Lyman edge likely due to obscuration by host galaxies\citep{Kriss:1997, Shull:2004}.  The central engines of active galactic nuclei (AGN)  have intermittent duty cycles, so the effects of previous AGN activity within an apparently dormant galactic environment may reduce, for a time, the local \ion{H}{1} density aiding LyC escape.  A quasar in close proximity to star-forming galaxies could produce a similar effect.  Such considerations become increasingly important as the MIB budget moves from one dominated by star-forming galaxies near $z \sim 6$ to one with a quasar contribution that peaks at $z \sim 2$ \citep{Cowie:2008}. Wide-field observation of $f_e$ from objects in extended cluster environments can be used to map out the relative contribution of galaxies, AGN and quasars to the MIB in the local universe and provide a means to assess its spatial uniformity in the redshift range $z \la 2.3$, which cannot be probed by the ratio of \ion{He}{2}/\ion{H}{1} \lya\ forest lines \citep{Shull:2004}.

\section{LyC and \lya\ Escape: Environments, Analogs and  Proxy Prospects}


Reionization appears to require LyC leakage from galaxies with $f_e \sim 0.1$, but how LyC and \lya\ escape from galaxies is somewhat mysterious.  Most star-forming galaxies have mean \ion{H}{1} columns greater than damped \lya\ systems (DLA).
The optical depth at the Lyman edge for DLAs is $\tau_{\lambda<912} > N_{HI}$ 6.3 x 10$^{-18}$ $(\lambda/912)^3$ = 1260 $(\lambda/912)^3$, while at the line core of \lya\ the optical depth is, $\tau_{Ly\alpha}$ = $N_{HI}$ 6.3 x 10$^{-14}$ = 1.26 $\times$ 10$^{7}$ (for $V_{dop}$ = 12 km s$^{-1}$).   Escape from such large mean optical depths requires that the interstellar medium (ISM) be highly inhomogeneous.  The escape of \lya\ ($f_{\alpha}$) is aided by velocity gradients and the presence of multi-phase media \citep{Dijkstra:2006, Hansen:2006, Verhamme:2006, Neufeld:1991}. Similarly, LyC escape is thought to result from galaxy porosity, low neutral density, high ionization voids, chimneys created by supernovae or the integrated winds from stellar clusters \citep{Fujita:2003},
 
Exploring the possibility of a proxy relationship between $f_{e}$  and  $f_{\alpha}$ will be extremely important in the coming decade as JWST seeks to identify the source(s) responsible for initiating and sustaining reionization. The brightness of \lya\ emission from the first objects is expected to be much easier to detect than their rest frame UV continuum.  Consequently, JWST will probably have to rely upon observations of \lya\ escape as a proxy for LyC escape. Unfortunately there is no guarantee that such a proxy relationship exists, because escaping \lya\ photons are created by recombining electrons freed by the LyC photons that do not escape \citep{Stiavelli:2004} ([\lya\ $\approx (2/3)Qf_{\alpha}(1-f_e)exp{(-\tau)}$]).  It is  essential to test the proxy hypothesis at $ z \lesssim$ 3 by obtaining simultaneous observations of LyC and \lya.   


\section{LyC escape detections and the advantages of spectroscopy}

\citet{Siana:2010} have provided the most current summary of detection efforts to date.  In short they have returned mixed, but tantalizing  results, that hint at a trend for $f_e$ falling towards low-$z$.  Spectroscopic observations hold an advantage over broad band imaging by providing the means to quantify ISM and IGM attenuation by \ion{H}{1} using Lyman series absorption and investigating \lya\ escape processes.  Moreover, they provide the means to detect spectroscopically, absorption features from species with wavelengths shortward of the Lyman edge, as was reported by \citet{Bogosavljevic:2009} who found evidence for  the \ion{O}{1} $\lambda$ 877 in a stack of 13 spectra of $z =$ 3 galaxies.  A quantitative assessment of the evolving contribution of galaxies to the MIB will likely require spectroscopic surveys over wide angular fields to acquire the large number of observations needed for establishing LyC luminosity function \citep{Deharveng:1997}.

\begin{figure}[t]
\includegraphics*[viewport=1in 1in 8.in 6in, scale=.5]{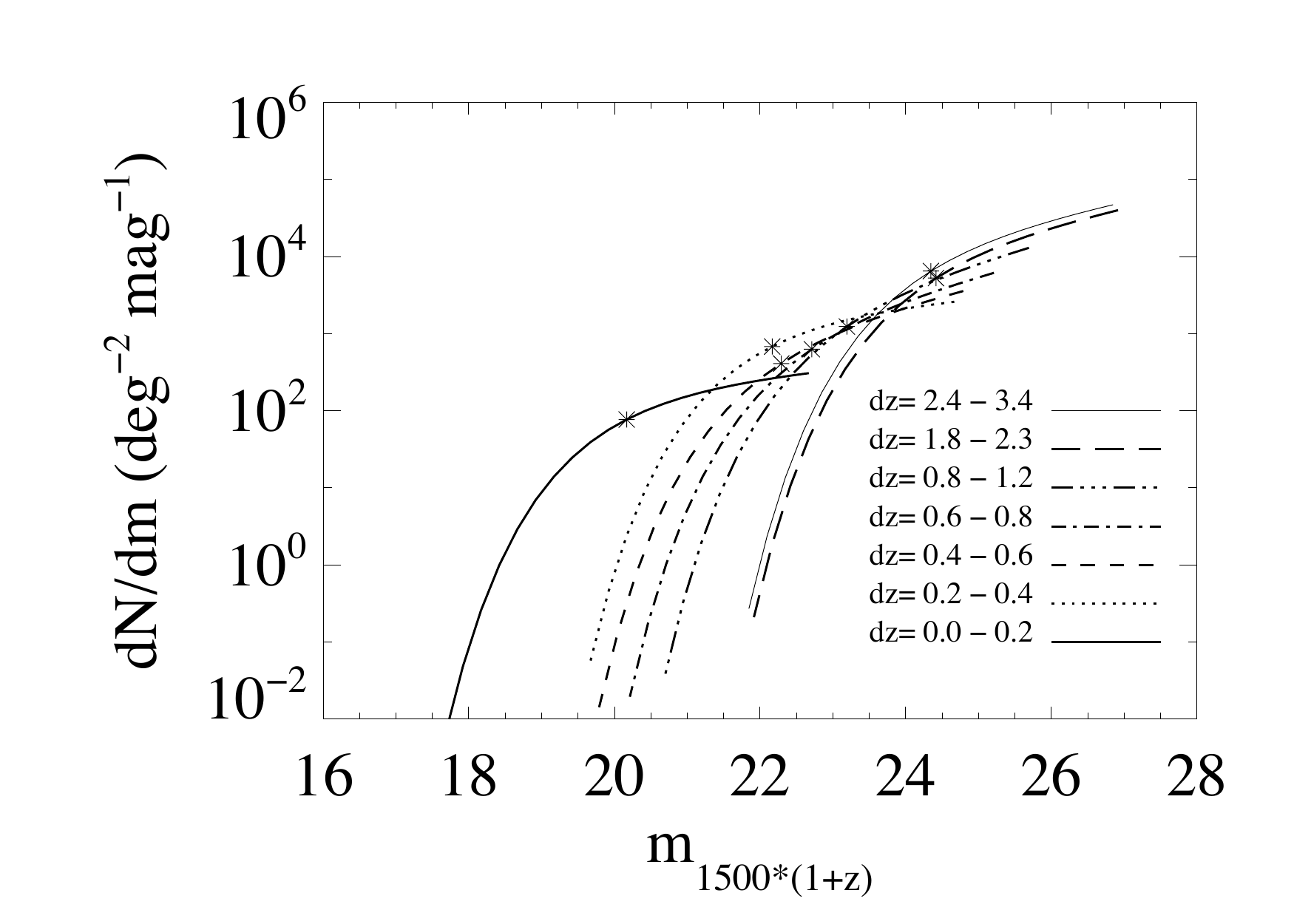} 
\parbox[b]{2.5in}{ \small {\textbf{\textsc{figure~4.} } } Surface densities as a function of observer's frame apparent  magnitude for galaxy populations with redshifts between 0 -- 0.2, 0.2 -- 0.4, 0.4 -- 0.6, 0.6 -- 0.8, 0.8 -- 1.2, 1.8 -- 2.3, 2.4 -- 3.4, estimated following Arnouts\cite{Arnouts:2005}.  There are 100s -- 10,000's of galaxies per square degree per magnitude with 24$ > m^*_{1500(1+z)} > $ 20 for each redshift interval. \label{lumfunback} \vspace*{.75in}}
\end{figure}
\setcounter{figure}{4}

\section{LyC detection requirements}

{\it GALEX} has shown there are thousands of far-UV emitting galaxies per square degree down to its limiting magnitude m$_{FUV} \approx$ 25.
\citet{McCandliss:2008} have suggested a wide field spectroscopy survey as an efficient way to search for LyC and \lya\ leakage.  Instrument requirements can be derived from the sensitivity required to detect LyC escaping star-forming galaxies over the redshift interval 0.02 $\lesssim z \lesssim$ 3.\footnote{The lower limit is set by the need to work at redshifts high enough to escape the H\,{\sc I} ``shadow'' of the Milky-Way.}  

To estimate the sensitivity requirement we use the surface density of UV rest frame emitting galaxies as a function of apparent magnitude (observer frame flux) and redshift, Figure~4, where the LFs from \citet{Arnouts:2005}, for the redshift intervals indicated in the caption, are shown.   Each asterisk marks the characteristic magnitude of the LF, appearing in the Schechter function.  

\begin{figure}
\includegraphics*[viewport= 0in 4.65in 8.5in 10.25in,scale=.6]{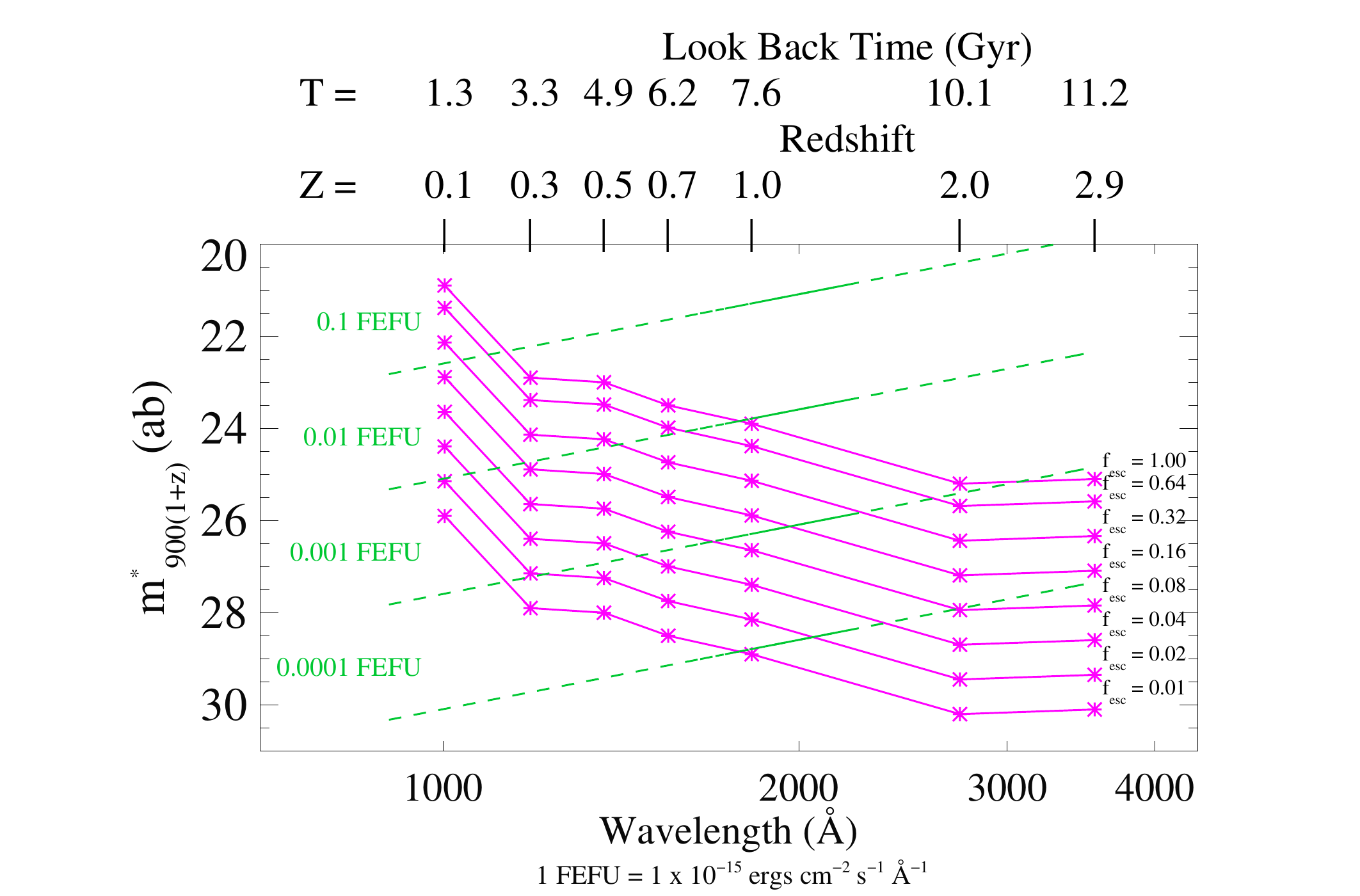}
\caption{The purple  asterisks show the characteristic apparent LyC magnitudes (ab) $m*_{900(1+z)}$  as a function of look back time, and in redshift and wavelength space, for different escape fractions.   Contours of constant flux units are overplotted as green dashes marked in FEFU fractions; the background limit for {\it FUSE}.   See \citep{McCandliss:2008} for details.}\label{back}
\end{figure}

We convert the 1500 \AA\ characteristic apparent magnitudes to LyC magnitudes using,
\begin{equation}
\label{eq2}
m^*_{900(1+z)} = m^{*}_{1500(1+z)}  + \delta m^{1500}_{900} + \delta m_{e},
\end{equation}
with $\delta m_{e} = -2.5 \log{f_{e}}$ and $\delta m^{1500}_{900} = 2.5\log{(f_{1500}/f_{900})}$. Starburst99 models\citep{Leitherer:1999} for  continuous star-formation, assuming solar metallicity, a Salpeter IMF and an upper mass cutoff of 100 M$_{\odot}$, yield $f_{1500}/f_{900} \approx$ 2. This ratio is insensitive to age with 1.5 $ \lesssim f_{1500}/f_{900} \lesssim$ 3 for ages 10 -- 900 Gyr.   The apparent magnitude in the LyC, as a function of $z$ for $f_{e}$ = 0.01, 0.02, 0.04, 0.08, 0.16, 0.132, 0.64 and 1 are displayed in Figure~\ref{back} as a series of purple connected asterisks.  Dashed green lines give the conversion from magnitude to flux units.  We find that $L_{uv}^*$ galaxies with $f_{e}$ = 1\% have LyC fluxes $<$ 10$^{-19}$ ergs cm$^{-2}$ s$^{-1}$ \AA$^{-1}$ at look back times between 7.6 and 11.2 Gyrs (1 $< z <$ 3).

\section{Central questions, prioritization and enabling technologies}

Understanding the mechanisms of reionization hinges on understanding how $f_e$ changes as a function of luminosity and redshift.  The answer will be important regardless of the outcome.  If star-forming galaxies are found with  $f_e \ga$ 0.1 then they become plausible sources of reionization.  If not, then new physics may be required to explain reionization \citep{Gnedin:2008}.

Four central questions of fundamental importance to the field can be addressed in the coming decade:

\begin{enumerate}

\item What are the relative contributions of quasars, AGN and star-forming galaxies to the MIB at $z \la$ 3.
\item What is the relationship between $f_e$ and the local and global parameters of metallicity, gas fraction, dust content, star formation history, mass, luminosity, redshift, over-density and quasar proximity.
\item Do low-$z$ analogs exist of the faint high-$z$ galaxies thought to be responsible for reionization?
\item Can the escape fraction of Ly$\alpha$ photons serve as a proxy for $f_e$?
\end{enumerate}
The answer to the last question is critical to the JWST key project seeking the source(s) of reionization.   

These questions can be most efficiently addressed using ultra-sensitive wide field spectroscopic surveys from space to probe the ionizing characteristics of galaxies as old as 11 Gyrs.  
Key mission-enabling technologies that will support the development of 0\fdg5 field-of-view, multi-object UV spectroscopy include microshutter arrays \citep{Kutyrev:2004}, high efficiency aberration corrected dual-order gratings \citep{McCandliss:2004}, large format detectors \citep{Fleming:2011} and GaN photocathodes \citep{Siegmund:2006}.  Future high-grasp UV telescopes sensitive enough to detect LyC leak will also easily detect the cosmic web of low surface brightness \lya\ emission, providing an unambiguous beacon for emerging complexity \citep{Furlanetto:2003,Latif:2011}.







\bibliographystyle{aipproc}   

\bibliography{RFILyCwhitepaper}

\IfFileExists{\jobname.bbl}{}
 {\typeout{}
  \typeout{******************************************}
  \typeout{** Please run "bibtex \jobname" to optain}
  \typeout{** the bibliography and then re-run LaTeX}
  \typeout{** twice to fix the references!}
  \typeout{******************************************}
  \typeout{}
 }

\end{document}